\documentclass[reprint,prb]{revtex4-2}
\usepackage{graphicx}
\usepackage{amssymb}
\usepackage{physics}
\usepackage{bbm}

\newcommand{\up}{\uparrow}
\newcommand{\dn}{\downarrow}

\begin{document}

\title{Connecting entanglement growth with local integrals of motion \\ in the disordered Fermi-Hubbard model}

\author{Ahad Nokhostin Helm}
\author{Brandon Leipner-Johns}
\author{Rachel Wortis}
\email{rwortis@trentu.ca}
\affiliation{Department of Physics \& Astronomy, Trent University, Peterborough, Ontario, K9L0G2, Canada}

\date{\today}

\begin{abstract}
Generically a quantum system initialized in an unentangled state will, under unitary dynamics, rapidly become entangled, a process closely related to information transport and to thermalization.  Disorder can suppress the growth of entanglement and result in memory of initial conditions.  In non-interacting systems this arises from localization of single-particle states, the occupancy of which is fixed by the initial condition.  In interacting systems similar localized conserved quantities persist, but with the added feature that they are coupled, resulting in entanglement growth which is distinct from both non-interacting localized systems and from generic ergodic systems.  The Fermi-Hubbard model has two degrees of freedom per site -- charge and spin -- and disorder may be present in both of these.  We study the growth of entanglement in two scenarios -- disorder in charge equal and unequal to that in spin, and determine the distinct contributions of charge and spin degrees of freedom by expanding the Hamiltonian in terms of a set of optimally localized conserved quantities with separate charge and spin character.  We find that coupling between charge and spin is significantly weaker than charge-charge and spin-spin coupling.  While this decoupling is present in all our results, it is only apparent when the strength of the disorder in the two sectors is different such that there is a separation between the characteristic timescales of the contributions to entanglement made by charge and by spin.
\end{abstract}

\maketitle 

\section{Introduction}
\label{sec:introduction}

Understanding the growth of entanglement is of fundamental interest because the entanglement of a system with its environment is the mechanism by which our classical world emerges from its quantum constituents.  
Moreover, controlling the growth of entanglement is key to applications in quantum computing, communications, and sensing, both to generate within a device the entanglement required for computation and simultaneously to limit entanglement with the environment.
Disorder influences the growth of entanglement in both non-interacting and interacting systems.
In a non-interacting system, disorder may give rise to Anderson localization:\cite{1958Anderson,1979Abrahams}  single-particle states are exponentially localized in space, resulting in insulating behaviour.  Without interactions, entanglement arises exclusively from particle motion, so localization caps the entanglement.  In an interacting system, entanglement between two subsystems can arise both from particle motion and from correlations in the configurations, and generically entanglement will be proportional to the system volume.\cite{1993Page,2005Calabrese}  With disorder and interactions, particle motion may be inhibited while correlations between configurations continue to develop, resulting in entanglement dynamics distinct from those in both non-interacting disordered systems and generic interacting systems.\cite{2008Znidaric,2012Bardarson,2013SerbynPRLJune}  This behaviour can be described in terms of a set of conserved quantities known as local integrals of motion (LIOMs).\cite{2013SerbynPRLJune,2014Huse}

Much of the existing work\cite{2015Nandkishore,2019Abanin,2021Abanin,2025Sierant} on disordered interacting systems has been on spin systems, or equivalently spinless fermions, in which there is a single degree of freedom per site--spin or charge.  
In these systems, the entanglement growth associated with interactions has been shown to be logarithmic in time up to a saturation value set by the system size.\cite{2008Znidaric,2012Bardarson}
This logarithmic dependence can be described in a weakly interacting picture by considering the coupling between two exponentially localized states separated by a distance $x$.\cite{2013SerbynPRLJune}  
The interaction energy, proportional to their overlap, is of order $e^{-x/\xi}$, and the timescale on which they become entangled, proportional to one over the interaction energy, is of order $e^{x/\xi}$.  
At time $t$, states a distance $\propto \xi \log (t)$ apart become entangled.  
While the saturation value grows with system size, it remains far below both the Page limit\cite{1993Page} and what is seen in delocalized systems.

The Anderson-Hubbard model, which incorporates inter-site hopping, on-site Coulomb repulsion, and disorder, is a particularly interesting area for further study because it has two degrees of freedom per site, both spin and charge, either or both of which may experience disorder.  In addition, this is a foundational model for some of the experimental systems of greatest current interest, including quantum simulators such as cold atoms in optical lattices as well as quantum materials such as transition metal oxides. 

The bulk of the theoretical work on the disordered Hubbard model has focused primarily on the question of whether localization occurs in the presence of disorder in just one of the two channels, charge 
\cite{2016Prelovsek,2018Mierzejewski, 2019Protopopov, 2019Leipner-Johns, 2023Thomson}
or spin.\cite{2018Yu}
There is strong consensus that continuous non-Abelian symmetry is not consistent with localization in all degrees of freedom due to the resulting resonances.
\cite{2016BarLevPRB,2016Potter,2017Protopopov,2018Kozarzewski,2019Protopopov, 2022Bonca}
Localization can nonetheless persist in special subspaces,\cite{2018Yu,2019Iadecola}
and evidence of localization in one channel while the other remains delocalized has been dubbed disorder-induced spin-charge separation.\cite{2023Thomson}
A number of mechanisms for breaking the SU(2) spin symmetry have been noted.  These include the addition of disorder in both charge and spin,\cite{2016Prelovsek,2018Mierzejewski,2022Bahovadinov}
a magnetic field gradient,\cite{2019Protopopov} spin-dependent hopping,\cite{2019Sroda} and disorder in interactions.\cite{2016BarLevPRB} 
Long-range interactions have also been argued to promote localization even while preserving spin symmetry.\cite{2020Pandey}
A number of these studies have calculated entanglement entropy, along with other measures, to support their arguments.\cite{2016BarLevPRB,2016Prelovsek,2018Zakrzewski,2020Pandey}
Only one\cite{2016Prelovsek} shows the entanglement versus time with disorder in both charge and spin (equal strength) and this includes only two orders of magnitude in time, limiting the conclusions that can be drawn.

Here we present a detailed examination of entanglement growth in the presence of disorder in both charge and spin, asking how the relative strengths of the disorder in these two channels influence the growth of entanglement.  
To provide insight into the observed growth we expand the Hamiltonian in terms of a set of localized conserved quantities with distinct charge and spin character\cite{2019Leipner-Johns} and examine the contribution of different terms in this expansion to the entanglement.
We find that when the locality of the integrals of motion is optimized, much of the physics is captured by just the first and second order terms in the expansion.
With strong disorder in both charge and spin, the entanglement growth follows a similar pattern to that seen in systems with only a single degree of freedom per site,\cite{2012Bardarson} namely an initial rapid rise followed by logarithmic growth.  
This reflects the exponential decay of couplings between LIOMs, but it hides a distinction:  coupling between charge and spin is much weaker than intra-species coupling.
When disorder in charge is strong while that in spin is weak, this decoupling becomes apparent, resulting in a distinct pattern of entanglement growth.

In Section \ref{sec:method} we describe the model and our approach, including the measure of entanglement and the construction of integrals of motion and expansion of the Hamiltonian in terms of them.  Section \ref{sec:results} presents our results, and Section \ref{sec:conclusion} provides some concluding remarks.

\section{Method}
\label{sec:method}

\begin{figure}[htbp]
   \centering
   \includegraphics[width=0.9\columnwidth]{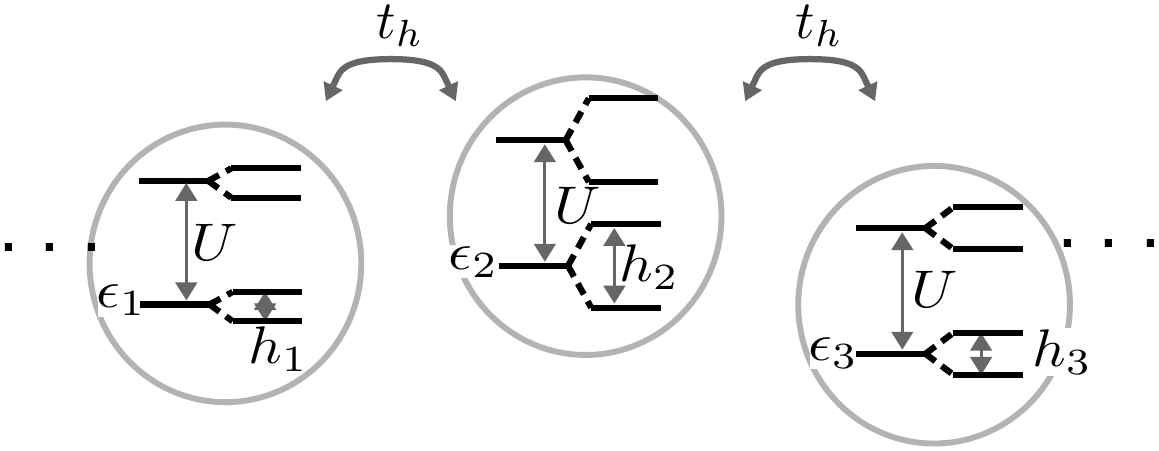} 
   \caption{A schematic diagram of the disordered Hubbard model showing three sites each with on-site Coulomb repulsion $U$, site potential $\epsilon_i$, and local magnetic field $h_i$, as well as inter-site hopping $t_h$.}
   \label{fig:model}
\end{figure}

We study the single-band Fermi-Hubbard model with nearest-neighbor hopping amplitude $t_h$, on-site Coulomb repulsion $U$, and disorder in both the site potentials and the local magnetic field, as represented in Fig.\ \ref{fig:model}.
\begin{eqnarray}
H &=& - t_h \sum_{\expval{i,j},\sigma} (c_{i\sigma}^{\dag} c_{j \sigma} 
	+ c_{j\sigma}^{\dag} c_{i\sigma})
	\nonumber \\ & &
	+ \sum_i ( U n_{i \up} n_{i\dn} + \epsilon_i d_i + h_i m_i)
\label{eqn:Hexact}
\end{eqnarray}
$c_{i\sigma}^{\dag}$ ($c_{i\sigma}$) are creation (annihilation) operators for a particle of spin $\sigma=\pm 1/2$ at site $i$.
$n_{i\sigma} \equiv c_{i\sigma}^{\dag} c_{i\sigma}$ is the corresponding number operator,
and $d_i\equiv n_{i\up}+n_{i\dn}$ and $m_i \equiv n_{i\up}-n_{i\dn}$ are the corresponding local charge and magnetization operators, respectively.
$\epsilon_i$ are independent and identically distributed random variables chosen from a flat distribution in the range $(-W_{ch},W_{ch})$.  Similarly for $h_i$ with the range $(-W_{sp},W_{sp})$.
We study one-dimensional systems with open boundary conditions.
The unit of energy is chosen to be the hopping amplitude, so $t_h=1$ and time has units of $\hbar/t_h$.
We focus here on $U=1$.  Preliminary results with larger values of $U$ do not show qualitatively different behaviour.
We consider two sets of disorder strengths:  one in which disorder is strong ($W>>U,t_h$) in both charge and spin, $W_{ch}=W_{sp}=16$, and one in which the disorder in spin is much less than that in charge, $W_{ch}=16$, $W_{sp}=0.1$.

As a measure of entanglement we use the bipartite von Neumann entanglement entropy between the left and right halves of the system.  This is calculated, for a pure state, at time $t$ from the density matrix of the full system $\rho(t) = \ket{\psi(t)} \bra{\psi(t)}$ traced over the left half of the system to obtain the reduced density matrix $\rho_R \equiv {\rm Tr}_L \rho$.  From the eigenvalues $p_i$ of $\rho_R$ the entanglement entropy is calculated: $S(t)=-\sum_i p_i \log p_i$, where a base two log is used.

\begin{figure}[htbp]
   \centering
   \includegraphics[width=\columnwidth]{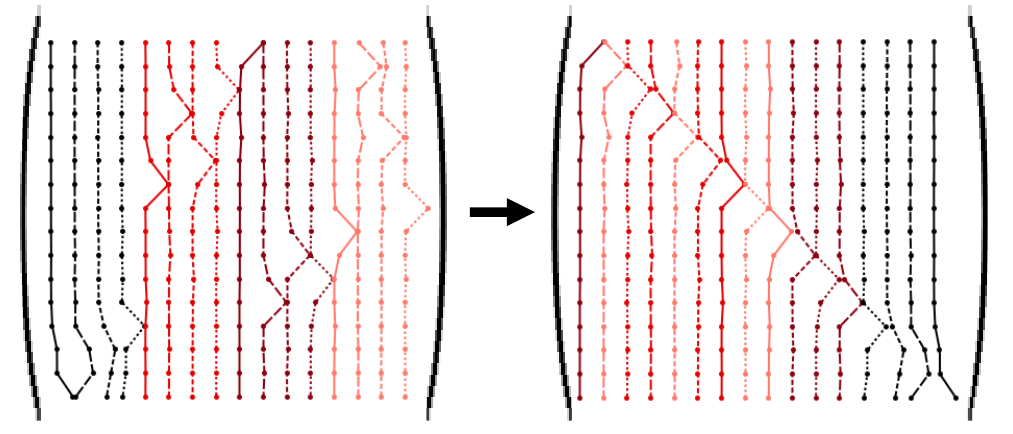} 
   \caption{
Graphical representation of a $Q$ matrix showing the re-ordering of the columns.
Each line represents one column.
The horizontal displacement of the $n$th dot in the $m$th line represents the magnitude of the element $Q_{nm}$.
On the left is the matrix before the Hungarian algorithm is applied, and on the right is the matrix after optimization.
}
   \label{fig:columns}
\end{figure}

To provide insight into the charge and spin contributions to the growth of entanglement, we identify optimally localized integrals of motion with charge and spin character\cite{2019Leipner-Johns} and express the Hamiltonian in terms of these.
To build the integrals of motion, we calculate all eigenvalues, $\{ E_n \}$, and eigenstates, $\{ \ket{E_n} \}$, using exact diagonalization, and construct the matrix $Q$ the columns of which are the eigenstates expressed in the Fock basis, $\{ \ket{m} \}$:  $Q_{mn}=\bra{m}\ket{E_n}$.
If an operator $A$ is diagonal in Fock space, the operator $QAQ^{\dag}$ is diagonal in the energy basis and hence conserved.
To make this integral of motion maximally local, we start with a local operator and endeavor to change it as little as possible by making $Q$ as close as possible to the identity.
Specifically, we choose the order of the columns in $Q$ such as to maximize its trace, as shown in Fig.\ \ref{fig:columns}.  This optimization task is achieved using the Hungarian algorithm.
The local operators we use are orthonormalized single-site charge and magnetization operators, ${\tilde d}_i \equiv \sqrt{2} (d_i -I)$ and ${\tilde m}_i \equiv \sqrt{2} m_i$, chosen to be orthogonal and normalized under the Frobenius inner product: $(A,B) = {\rm Tr} (A^{\dag} B)$. 
From these we construct charge and spin specific LIOMs, 
\begin{eqnarray}
\mathbbm{d}_i &\equiv& Q{\tilde d}_i Q^{\dag} \ \ {\rm and} \ \ \mathbbm{m}_i \ \equiv \ Q {\tilde m}_i Q^{\dag}. 
\label{eqn:unitarytrans}
\end{eqnarray}

Next we express the Hamiltonian in terms of these LIOMs.
This is achieved by projecting the Hamiltonian onto a complete set of operators composed of products of LIOMs.
Further details of the construction of this set are provided in Appendix \ref{app:set}.
\begin{eqnarray}
H &=& J_0 \mathbbm{I} 
		+ \sum_i J_i^c \mathbbm{d}_i + \sum_i J_i^s \mathbbm{m}_i \nonumber \\ & & 
		+ \sum_{i,j>i} J_{ij}^{cc} \mathbbm{d}_i \mathbbm{d}_j
		+ \sum_{i,j>i} J_{ij}^{ss} \mathbbm{m}_i \mathbbm{m}_j
		+ \sum_{i,j\ne i} J_{ij}^{cs} \mathbbm{d}_i \mathbbm{m}_j  \nonumber \\ & & 
		+ \sum_i J_i^o \mathbbm{o}_i
		+ ...
\label{eqn:Hliom} \\
{\rm where} & & J_i^c = (H,\mathbbm{d}_i), 
	\ J_{ij}^{cs} = (H,\mathbbm{d}_i \mathbbm{m}_j), \ {\rm etc} \nonumber
\end{eqnarray}
When all terms in this sum are included this is simply a rewriting of the original Hamiltonian in Eq.\ (\ref{eqn:Hexact}).
We study the dynamics of this full Hamiltonian (sometimes written $H_{exact}$ for emphasis) as well as the dynamics of several truncations, for example dropping all terms beyond 1st order ($H_{liom1}$) or all terms beyond 2nd order ($H_{liom2}$).

Results are averaged over disorder configurations and over initial conditions randomly selected from half-filled, zero-net-spin Fock states.
Eigenvalues and eigenvectors are accurate to at least 10 significant figures (based on floating point operations and comparisons between distinct eigensolvers), allowing confidence in evolution to long times.


\section{Results}
\label{sec:results}

In this section we present our results, starting first with the case of strong disorder in both charge and spin and then considering the case in which the spin disorder is much weaker than that in the charge.

\subsection{Equal disorder}

\begin{figure}[htbp]
\raggedright
   \includegraphics[width=0.75\columnwidth]{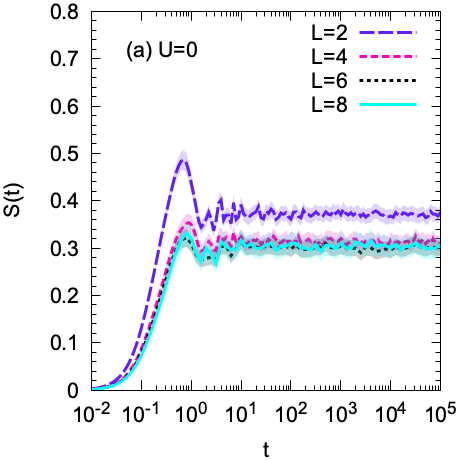} 
   \\
   \includegraphics[width=\columnwidth]{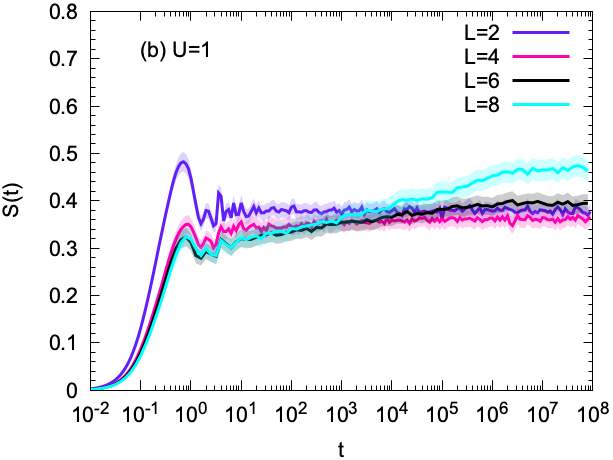} 
   \caption{Entanglement entropy $S(t)$ vs $\log(t)$ calculated using the full Hamiltonian $H_{exact}$ (a) non-interacting $U=0$ and (b) interacting $U=1$.  
The disorder strength is equal in charge and spin:  $W_{ch}=W_{sp}=16$.  System sizes $L=2,4,6,8$.
Results are averaged over 2000 disorder configurations and initial states for $L=2$, 1800 for $L=4$, 1600 for $L=6$, and 1000 ($U=0$) and 1500 ($U=1$) for $L=8$.
	}
   \label{fig:WeqHex}
\end{figure}

Fig.\ \ref{fig:WeqHex} shows the entanglement entropy obtained from the full Hamiltonian with strong disorder in both charge and spin.
For $U=0$, the entanglement has a rapid initial rise to a peak at a time of order one, followed by a dip and then saturation.
This behaviour is consistent with results in single-component systems.
The reason for this structure can be understood by considering a single particle initially placed at one site.  
Without interactions, up and down spins are independent and Anderson localized.
The particle's wavefunction is a superposition of the set of localized single-particle eigenstates which have nonzero overlap with the site on which it is initially placed.
As time evolves the wavefunction amplitude shifts outward into these localized eigenstates, increasing entanglement.  
When averaged over disorder configurations, this produces a maximum expansion and a slight contraction (the peak-dip structure) before settling to a constant (the saturation of entanglement) at a value set by the localization length, $\xi$.
The behaviour is qualitatively similar for all system sizes, with the location of the peak being independent of system size, while the peak height and saturation value are slightly elevated for the smallest systems, in which the system size is of order the localization length. 

With interactions, the same initial rise, peak, and dip are seen, reflecting very similar dynamics to the non-interacting case but with LIOMs taking the place of the Anderson localized single-particle states.
Here, however, the LIOMs are coupled, allowing for further growth as described in the introduction.
In the $L=2$ and $L=4$ systems, in which the length scale of the LIOMs is comparable to the system size, the initial peak is higher, as seen in the non-interacting case, and the size-dependent saturation value of the entanglement is low, such that further growth in the entanglement is negligible.
In the $L=6$ and $L=8$ systems, however, logarithmic time dependence is seen, starting around $t=10$ and extending to roughly $t=10^3$ for $L=6$ and several orders of magnitude further for $L=8$.  

\begin{figure}[htbp]
   \centering
   \includegraphics[width=\columnwidth]{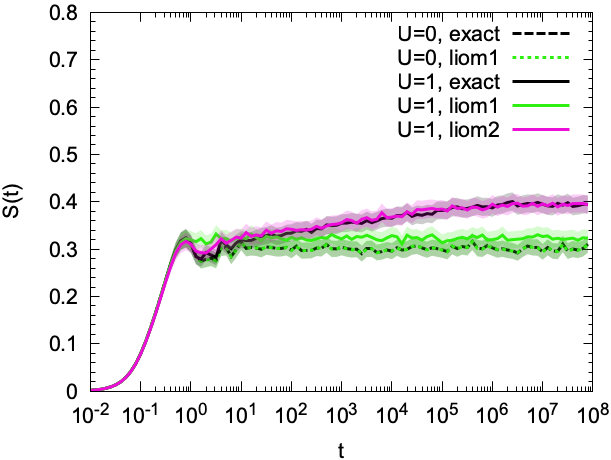} 
   \caption{Entanglement entropy $S(t)$ vs $\log(t)$ calculated using the full Hamiltonian $H_{exact}$ (black), the LIOM Hamiltonian keeping only up to 1st-order terms $H_{liom1}$ (green), and only up to 2nd-order terms $H_{liom2}$ (pink).  
Non-interacting results are shown dashed and interacting with solid lines.  
The disorder strength is equal in charge and spin:  $W_{ch}=W_{sp}=16$.  System size $L=6$.
Results in each case are averaged over (the same) 1600 disorder configurations and initial states.
LIOMs are optimized using $n_{HA}=7$.
	}
   \label{fig:WeqHexliom1liom2}
\end{figure}

Fig.\ \ref{fig:WeqHexliom1liom2} compares these exact results with the entanglement obtained from the Hamiltonian in Eq.\ (\ref{eqn:Hliom}) including terms only up to first order ($H_{liom1}$) and only up to second order ($H_{liom2}$).
Without interactions, the entanglement produced by $H_{liom1}$ is identical to the exact result:  The LIOMs alone, with no couplings, describe the Anderson localized physics.
With interactions, the entanglement produced by $H_{liom1}$ follows the exact result at early times but then saturates, looking similar to the non-interacting case with a slightly higher saturation value consistent with an increased localization length.
Meanwhile, the entanglement produced by $H_{liom2}$ is essentially the same as the exact result at all times, demonstrating the efficiency of the LIOM description.

\begin{figure}[htbp]
   \centering
   \includegraphics[width=\columnwidth]{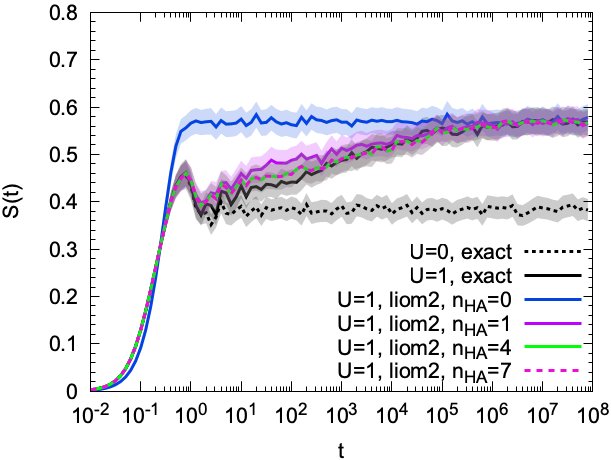} 
   \caption{Entanglement entropy $S(t)$ vs $\log(t)$ calculated using the full Hamiltonian $H_{exact}$ (black) and the LIOM Hamiltonian keeping only up to 2nd-order terms $H_{liom2}$ where the LIOM are generated using a $Q$ matrix with columns in random order (blue), optimized to one digit (purple), to four digits (green), and to seven digits (pink).
Here $W_{ch}=W_{sp}=16$, $L=6$, and results are averaged over (the same) 1000 disorder configurations all with the initial condition $\ket{\uparrow\downarrow\uparrow\downarrow\uparrow\downarrow}$.
	}
   \label{fig:Weqoptimize}
\end{figure}

This match between the exact results and those of the LIOM Hamiltonian truncated at second order depends on the optimization of the locality of the LIOMs.
To highlight this,  Fig.\ \ref{fig:Weqoptimize} shows the entanglement produced by $H_{liom2}$ for four different levels of optimization.  
Specifically, when running the Hungarian algorithm, we keep different numbers of digits $n_{HA}$ in the elements of the matrix of eigenvectors $Q$.  
$n_{HA}=0$ means no optimization has been done: the columns in $Q$ are in random order.
The entanglement growth of the 2nd-order truncated LIOM Hamiltonian in this case bears no resemblance to the exact result.
Optimizing with the Hungarian algorithm keeping just one digit of each matrix element is already significantly better, and beyond four digits there is no further change.
$n_{HA}=7$ is for all other results.

\begin{figure}[htbp]
   \centering
\includegraphics[width=\columnwidth]{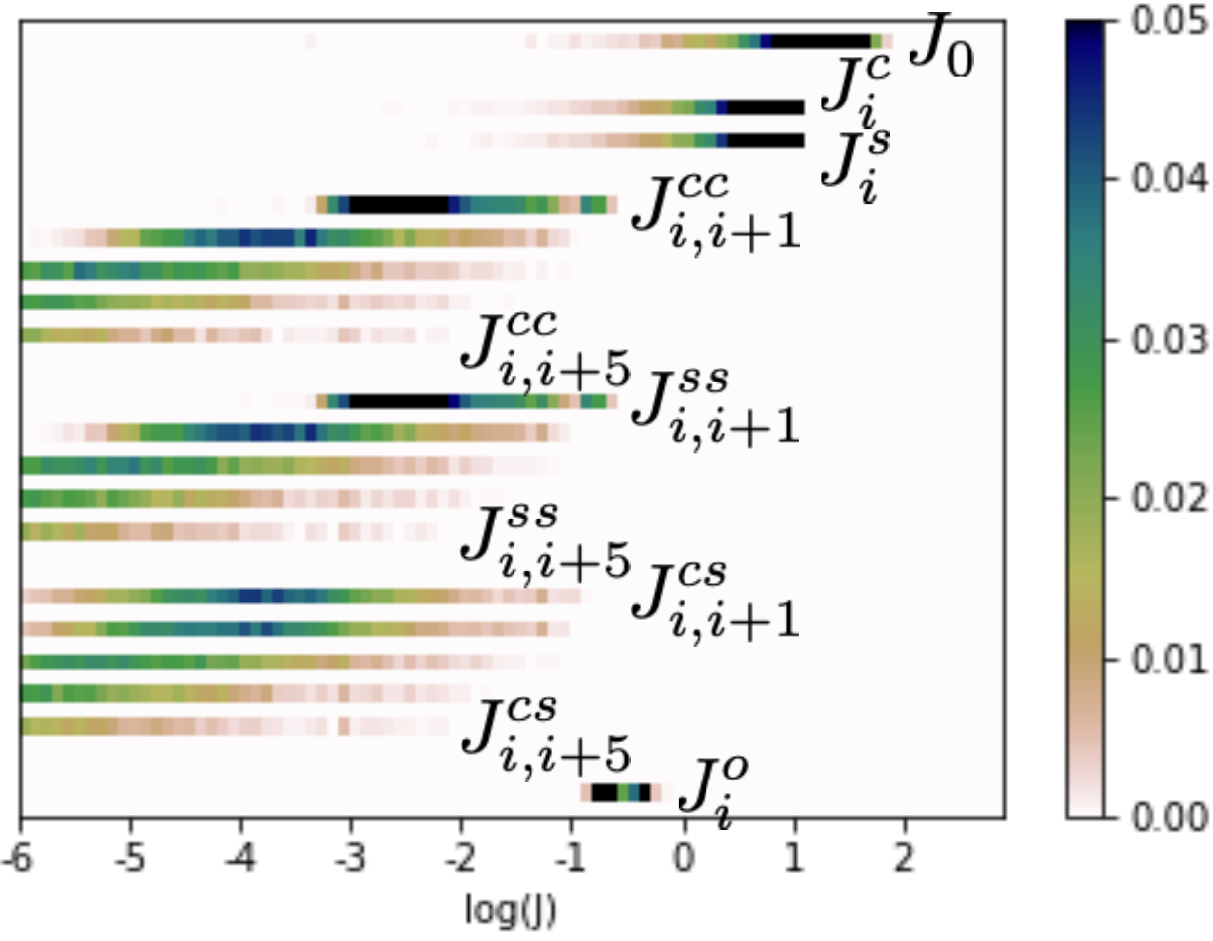} 
   \caption{Distributions of each category of coefficient up to second order:  The colour scale indicates the magnitude of the distribution.  The horizontal axis is the coupling constant magnitude on a log scale.  Each row corresponds to a different category of coefficient.  From top to bottom:  0th-order ($J_0$), 1st-order charge ($J_i^c$), 1st-order spin ($J_i^s$), 2nd-order charge-charge for nearest neighbors ($J_{i,i+1}^{cc}$) and so on to fifth neighbor, 2nd-order spin-spin ($J_{i,i+n}^{ss}$), 2nd-order charge-spin ($J_{i,i+n}^{cs}$), and 2nd-order onsite-terms ($J_i^o$).
Here $W_{ch}=W_{sp}=16$, $U=1$, $L=6$, $n_{HA}=7$, and 1600 disorder configurations and initial states are used.
	}
   \label{fig:Weqcoeff}
\end{figure}

Fig.\ \ref{fig:Weqcoeff} shows the distribution of the magnitudes of the coefficients in the Hamiltonian Eq.\ (\ref{eqn:Hliom}) for 1st-order and 2nd-order terms.  
With equal disorder in charge and spin, the distributions of 1st-order charge and spin coefficients are the same.
Similarly charge-charge and spin-spin coefficients have the same distributions, decaying exponentially with the distance between the generating operators of the corresponding LIOMs.
The coefficients of charge-spin terms also show exponential decay with distance for distances greater than two.
However, the nearest-neighbor coefficients are the same as the next-nearest-neighbor coefficients, and strikingly an order of magnitude smaller than the nearest-neighbor charge-charge and spin-spin coefficients.  
Prior work on Fermi-Hubbard systems with disorder in only one channel noted the apparent decoupling of charge and spin.\cite{2016Prelovsek,2018Mierzejewski,2018Yu,2023Thomson}
The suppressed inter-species coupling relative to the intra-species values demonstrates that aspects of this decoupling persist even when disorder is present in both charge and spin.
Nonetheless, when the disorder strength is the same in both charge and spin, $S(t)$ looks similar to that seen in single-species systems and does not show signatures of this decoupling of charge and spin.

\subsection{Unequal disorder}

\begin{figure}[htbp]
\raggedright
   \includegraphics[width=0.75\columnwidth]{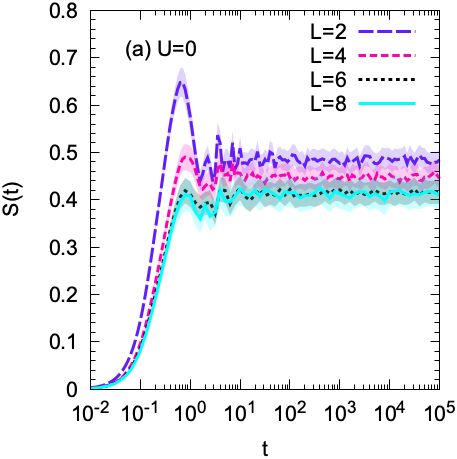} 
   \\
   \includegraphics[width=\columnwidth]{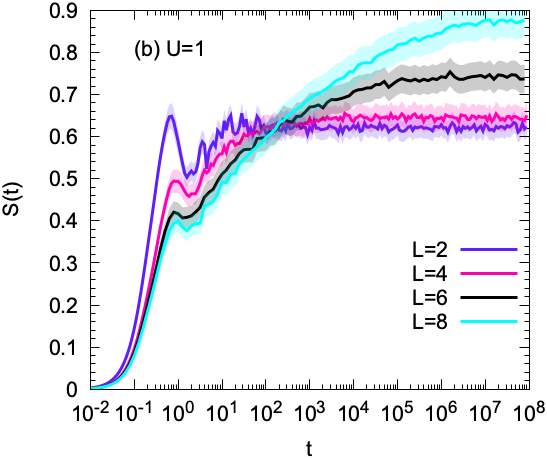} 
   \caption{Entanglement entropy $S(t)$ vs $\log(t)$ calculated using the full Hamiltonian $H_{exact}$ (a) non-interacting $U=0$ and (b) interacting $U=1$.  
The disorder strength is greater in charge than spin:  $W_{ch}=16,\ W_{sp}=0.1$.  System sizes $L=2,4,6,8$.
Results are averaged over 2000 disorder configurations and initial states for $L=2$, 1800 for $L=4$, 1600 for $L=6$, and 970 ($U=0$) and 1500 ($U=1$) for $L=8$.
	}
   \label{fig:WneHex}
\end{figure}

We now move on to the case in which disorder in spin is much less than that in charge.
Fig.\ \ref{fig:WneHex} shows the entanglement growth from the full Hamiltonian.  
Without interactions, the results are very similar to those seen with equal disorder in charge and spin.
The slightly higher saturation value, relative to Fig.\ \ref{fig:WeqHex}, is consistent with a lower level of disorder and a correspondingly longer localization length.

With interactions (solid), the early-time rise, peak, and dip are also essentially the same.
Regarding size dependence, as in the case of equal disorder, after the peak structure there is a region of parallel growth, with smaller systems peeling off as their saturation value is approached.
However, the continued growth is markedly different from the $\log(t)$ dependence seen with equal disorder and in single-component systems.
Comparison with the entanglement produced by truncated versions of the Hamiltonian Eq.\ (\ref{eqn:Hliom}) shed light on this distinct time dependence.

\begin{figure}[htbp]
   \centering
   \includegraphics[width=\columnwidth]{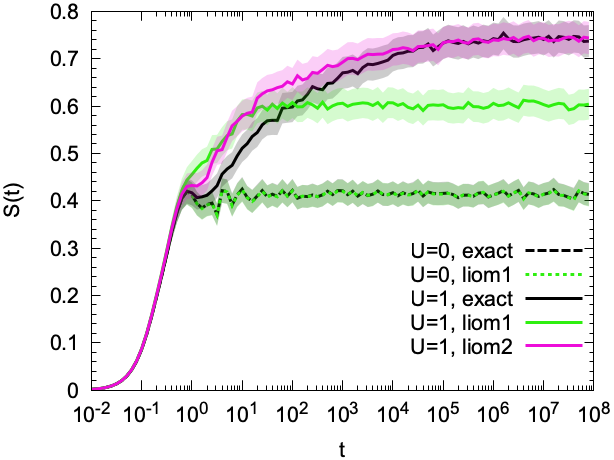} 
   \caption{Entanglement entropy $S(t)$ vs $\log(t)$ calculated using the full Hamiltonian $H_{exact}$ (black), the LIOM Hamiltonian keeping only up to 1st-order terms $H_{liom1}$ (green), and only up to 2nd-order terms $H_{liom2}$ (pink).  
Non-interacting results are shown dashed and interacting with solid lines.  
The disorder strength is greater in charge than spin:  $W_{ch}=16,\ W_{sp}=0.1$.  System size $L=6$.
Results in each case are averaged over (the same) 1600 disorder configurations and initial states.
LIOMs are optimized using $n_{HA}=7$.
	}
   \label{fig:WneHexliom1liom2}
\end{figure}

\begin{figure}[htbp]
   \centering
   \includegraphics[width=\columnwidth]{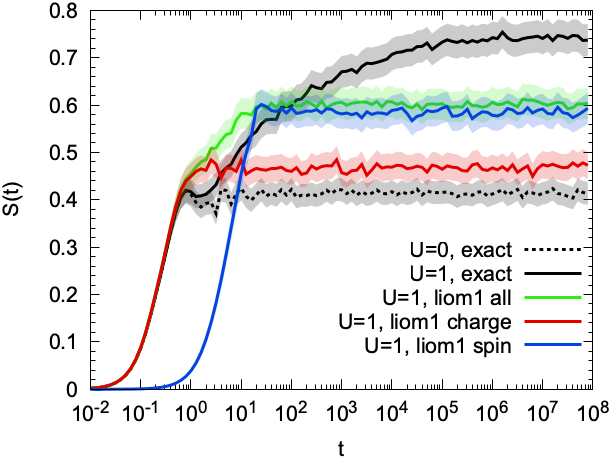} 
  \caption{Entanglement entropy $S(t)$ vs $\log(t)$ calculated using the full Hamiltonian $H_{exact}$, the LIOM Hamiltonian keeping only up to 1st-order terms $H_{liom1}$ (green), only 1st-order in charge (but not spin) $H_{liom1c}$ (red), and only 1st-order in spin (but not charge) $H_{liom1s}$ (blue).
The disorder strength is greater in charge than spin:  $W_{ch}=16,\ W_{sp}=0.1$.  System size $L=6$.
Results in each case are averaged over (the same) 1600 disorder configurations and initial states.
LIOMs are optimized using $n_{HA}=7$. 
	}
   \label{fig:WneHexliom1cs}
\end{figure}

Fig.\ \ref{fig:WneHexliom1liom2} shows the comparison between the exact entanglement and that obtained from the 1st-order and 2nd-order truncations of Eq.\ (\ref{eqn:Hliom}).  
Without interactions, as for equal disorder, just the LIOMs with no couplings capture the Anderson localized physics.
With interactions, there are differences.
First, the entanglement from the 1st-order terms has two distinct rises before saturation, the first similar to that seen in the non-interacting case and the second distinct.  
The reason for the two stage rise is apparent when one looks separately at the 1st-order charge and spin terms, as shown in Fig.\ \ref{fig:WneHexliom1cs}.  The first stage of the rise is coming from the charge, while the timescale of the entanglement growth in the spin is longer.

Returning to Fig.\ \ref{fig:WneHexliom1liom2}, the entanglement growth from the LIOM Hamiltonian truncated at second order is similar to the exact result in the early-time rise and in the saturation value.  
However, the continued rise after the peak is both earlier and slightly faster than in the exact case.
This can be understood as follows:
The higher-order terms in the Hamiltonian, not included in $H_{liom2}$, have progressively smaller coefficients.
Smaller coefficients correspond to longer timescales.  
Removing these terms shifts the entanglement growth to shorter timescales.

\begin{figure}[htbp]
   \centering
   \includegraphics[width=\columnwidth]{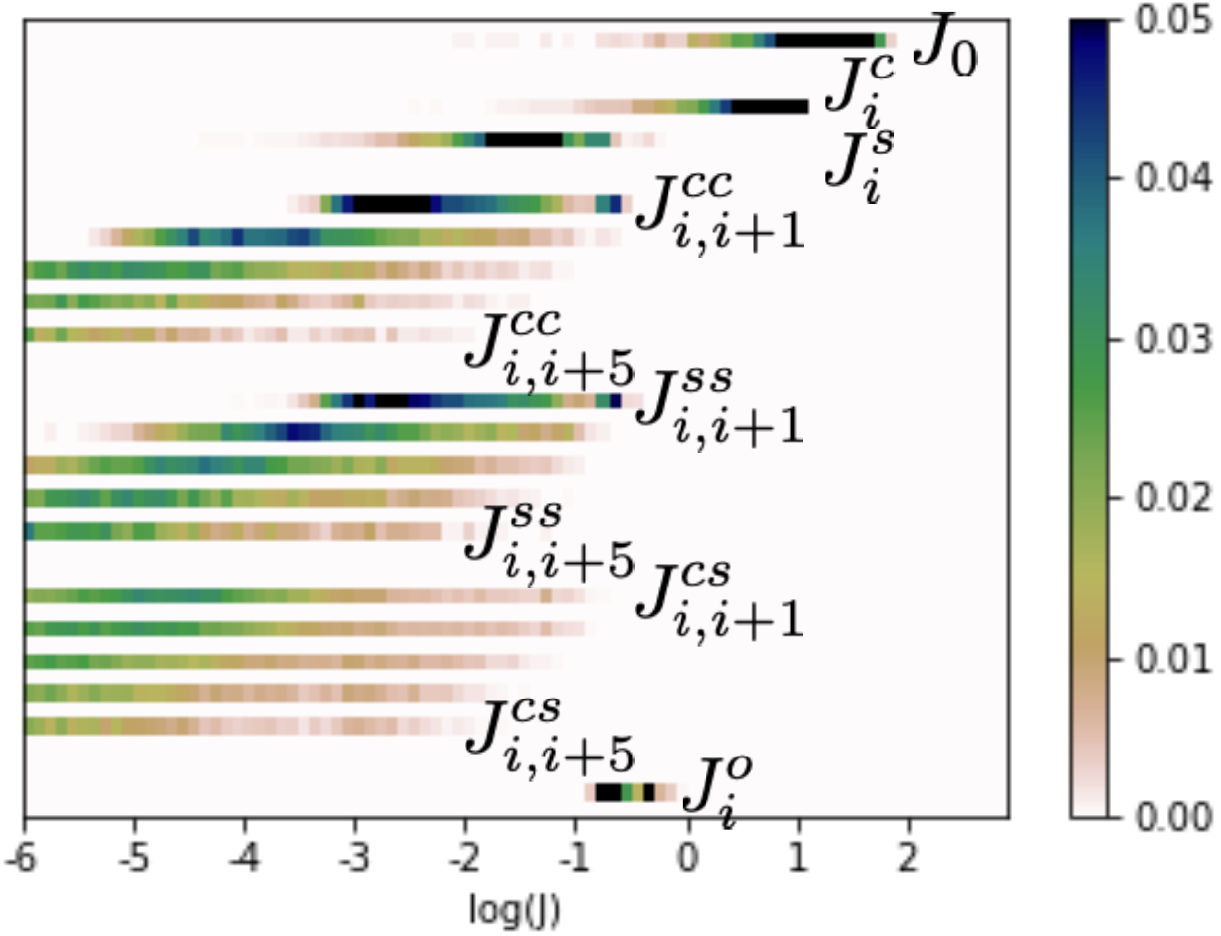} 
   \caption{Distributions of each category of coefficient up to second order:  The colour scale indicates the magnitude of the distribution.  The horizontal axis is the coupling constant magnitude on a log scale.  Each row corresponds to a different category of coefficient.  From top to bottom:  0th-order ($J_0$), 1st-order charge ($J_i^c$), 1st-order spin ($J_i^s$), 2nd-order charge-charge for nearest neighbors ($J_{i,i+1}^{cc}$) and so on to fifth neighbor, 2nd-order spin-spin ($J_{i,i+n}^{ss}$), 2nd-order charge-spin ($J_{i,i+n}^{cs}$), and 2nd-order onsite-terms ($J_i^o$).
Here $W_{ch}=16$, $W_{sp}=0.1$, $U=1$, $L=6$, $n_{HA}=7$, and 1600 disorder configurations are used.
	}
   \label{fig:Wnecoeff}
\end{figure}

Fig.\ \ref{fig:Wnecoeff} shows the first and second order coefficients in detail.  
Relative to the case of equal disorder strength, reducing the spin disorder results in a reduction in the coefficients of the 1st-order spin terms, corresponding to the delayed growth of the $H_{liom1s}$ data shown in Fig.\ \ref{fig:WneHexliom1cs}.
As for the 2nd-order terms, reduced spin disorder results in slower decay of the coefficients with distance, with the largest effect seen in the spin-spin terms.
The order of magnitude difference in timescales associated with intra-species and inter-species coupling persists.
Indeed, it is due to this decoupling that the separation of 1st-order charge and spin timescales seen in Fig.\ \ref{fig:WneHexliom1cs} remain visible in the full time evolution. 

\begin{figure}[htbp]
   \centering
   \includegraphics[width=\columnwidth]{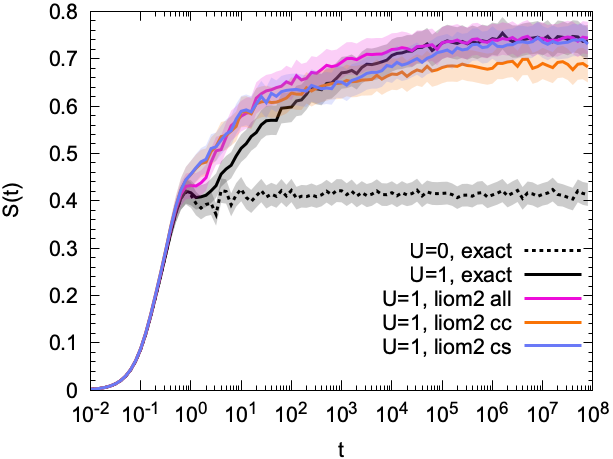} 
   \caption{Entanglement entropy $S(t)$ vs $\log(t)$ calculated using the full Hamiltonian $H_{exact}$, the LIOM Hamiltonian keeping only up to 2nd-order terms $H_{liom1}$ (pink), only 2nd-order charge-charge $H_{liom2cc}$ (orange), and 2nd-order charge-spin $H_{liom2cs}$ (violet).
The disorder strength is greater in charge than spin:  $W_{ch}=16,\ W_{sp}=0.1$.  System size $L=6$.
Results in each case are averaged over (the same) 1600 disorder configurations and initial states.
LIOMs are optimized using $n_{HA}=7$.
	}
   \label{fig:WneHexliom2cs}
\end{figure}

Fig.\ \ref{fig:WneHexliom2cs} shows the contributions of different 2nd-order terms to the entanglement growth.  
Comparing the entanglement from Eq.\ (\ref{eqn:Hliom}) with 1st-order terms plus only the charge-charge terms ($H_{liom2cc}$) with the corresponding result with 1st-order plus only the charge-spin terms ($H_{liom2cs}$), one can see that the contribution to entanglement of the 2nd-order charge-spin coupling persists to longer times and generates more entanglement than does the 2nd-order charge-charge coupling.
Also, it appears that the 2nd-order onsite terms are at least partially responsible for the dip after the peak, absent in the $H_{liom2cc}$ and $H_{liom2cs}$ results.

In this case of unequal disorder, the entanglement growth is not proportional to $\log(t)$.  
This is true even for $t > 10^2$ when the 1st-order contributions have saturated, and even in the $L=8$ case shown in Fig.\ \ref{fig:WneHex}(b).  
Some perspective can be provided by returning to the argument\cite{2013SerbynPRLJune} sketched in the introduction.
When there are distinct localization lengths, the interaction energy associated with the two localized states separated by a distance $x$ is not a single exponential but a sum of exponentials.
As a result, the timescale associated with entanglement on this length scale is not a single exponential but instead the inverse of a sum of exponentials,
$t \sim (e^{-x/\xi_c}+e^{-x/\xi_s})^{-1}$, and $x$ is no longer simply proportional to $\log(t)$.

\section{Conclusion}
\label{sec:conclusion}

We have examined the growth of entanglement in the Fermi-Hubbard model with disorder in both charge and spin.
In particular, we have explored the contribution of charge and spin sectors to the growth of entanglement.

When the strength of disorder is the same in charge and in spin, the growth of entanglement shows the characteristic $\log(t)$ behaviour seen in single-component systems.
To understand this, we construct the LIOM Hamiltonian and use truncated versions to calculate the entanglement growth.
When the locality of the LIOMs is optimized, keeping only up to 2nd-order terms in the LIOM Hamiltonian captures the exact behaviour well, an encouraging result for the potential of approximate approaches.  
By examining the coefficients in the LIOM Hamiltonian, we see the exponential decay with distance that gives rise to this $\log(t)$ behaviour.
With equal disorder in charge and spin, the rates of decay are the same, such that the very weak coupling between charge and spin is not apparent.

Different behaviour is seen when the strength of disorder is greater in charge than in spin.
The early-time growth arising from 1st-order terms in the LIOM Hamiltonian has two distinct phases, the first dominated by the charge and the second the spin.  
The fact that these contributions can be distinguished is made possible by the very weak coupling between charge and spin.
Similarly, at longer times, the growth is not logarithmic, reflecting distinct 2nd-order couplings which decay with different characteristic lengths.

For theoretical work some areas awaiting further exploration are examination of the dependence on initial conditions, as well as possibly distinct behaviour between strongly and weakly interacting regimes.
Further developments can also be hoped for on the experimental front where  ingenious work has shed light on both entanglement\cite{2015Islam,2019Lukin,2023Leonard} and coupling between LIOMs\cite{2022Chiaro}, although so far exclusively in bosonic systems.  




\section*{Acknowledgments}
We acknowledge support by the National Science and Engineering Research Council (NSERC) of Canada.
This work was made possible by the facilities of the
Shared Hierarchical Academic Research Computing Network
(SHARCNET). 
We also acknowledge helpful conversations with Malcolm Kennett, Bill Atkinson, and Rudolf R\"omer.

\appendix

\section{Complete set of operators}
\label{app:set}

We construct a set of operators which is complete in the sense that any operator diagonal in the energy basis may be expressed as a linear combination of elements of this set.
The set has a one-to-one correspondence with a set of operators in terms of which any operator diagonal in the Fock basis may be expanded.
We work in this more familiar context before transforming, by $Q$, to the energy basis.

Considering first just a single site, $i$.  The Hilbert space dimension is four, hence four operators are required to span the space of diagonal operators. 
The normalized identity operator ${\tilde I}_i^{(1)}=I/2$, the charge operator ${\tilde d}_i^{(1)}$, and the spin operator ${\tilde m}_i^{(1)}$ provide three of these, with diagonal elements $(1,1,1,1)/2$, $(-1,0,0,+1)/\sqrt{2}$, and $(0,+1,-1,0)/\sqrt{2}$, respectively.  
The superscripts emphasize that these act in the single-site basis.
To complete the set we define the operator ${\tilde o}_i^{(1)} \equiv ({\tilde d}_i^{(1)})^2 - ({\tilde m}_i^{(1)})^2$ with diagonal elements $(+1,-1,-1,+1)/2$.  
These four operators commute because they are all diagonal and are orthonormal by construction.

Returning to the case of $L$ sites, there are four single-site operators for each site $i$:  ${\tilde I}_i^{(1)}$, ${\tilde d}_i^{(1)}$, ${\tilde m}_i^{(1)}$, and ${\tilde o}_i^{(1)}$.
Choosing one of these four for each site and taking an outer product, we construct $4^{L}$ $L$-site operators, hence spanning the space of diagonal operators.
We define the order of our $L$-site operators by the number of charge and magnetization operators they contain.
For example, there is 
a single 0th-order operator 
${\tilde I} \equiv {\tilde I}_1^{(1)} \otimes {\tilde I}_2^{(1)} ... \otimes {\tilde I}_L^{(1)}$;
$2L$ 1st-order terms, such as
${\tilde d}_i \equiv {\tilde I}_1^{(1)}  ... \otimes {\tilde d}_i^{(1)} ... \otimes {\tilde I}_L^{(1)}$
or ${\tilde m}_i \equiv {\tilde I}_1^{(1)}  ... \otimes {\tilde m}_i^{(1)} ... \otimes {\tilde I}_L^{(1)}$;
$2L^2-L$ 2nd-order terms, etc.
A complete list for the case $L=2$ is shown in Table \ref{table:set}.
Note that the operator ${\tilde o}_i$ is second order, while acting only on a single site.

\begin{table}
\begin{tabular}{|l | l |} \hline
order & operators for two-site system \\ \hline
0th & ${\tilde I}$ \\
1st & ${\tilde d}_1$, ${\tilde d}_2$, ${\tilde m}_1$, ${\tilde m}_2$ \\
2nd & ${\tilde d}_1 {\tilde d}_2$, ${\tilde d}_1 {\tilde m}_2$, 
	${\tilde m}_1 {\tilde d}_2$, ${\tilde m}_1 {\tilde m}_2$, 
	${\tilde o}_1$, ${\tilde o}_2$ \\
3rd & ${\tilde d}_1 {\tilde o}_2$, ${\tilde o}_1 {\tilde d}_2$,
	${\tilde m}_1 {\tilde o}_2$, ${\tilde o}_1 {\tilde m}_2$ \\
4th & ${\tilde o}_1 {\tilde o}_2$ \\ \hline
\end{tabular}
\caption{\label{table:set}Elements of the complete set of operators spanning the space of operators diagonal in the Fock basis for a two-site system, grouped by order.}
\end{table}

This complete orthonormal set of operators in Fock space has a corresponding set in energy space generated by the unitary transformation in Eq.\ (\ref{eqn:unitarytrans}).
For example, 
\begin{eqnarray}
Q {\tilde o}_1 Q^{\dag} &=& \mathbbm{o}_1, \ {\rm and} \\
Q {\tilde d}_1 {\tilde m}_2 Q^{\dag} &=& \mathbbm{d}_1 \mathbbm{m}_2, \ {\rm etc}.
\end{eqnarray}
To write the Hamiltonian in terms of this set, the Hamiltonian is projected onto each of these operators to obtain the corresponding coefficient.


\begin{thebibliography}{32}%
\makeatletter
\providecommand \@ifxundefined [1]{%
 \@ifx{#1\undefined}
}%
\providecommand \@ifnum [1]{%
 \ifnum #1\expandafter \@firstoftwo
 \else \expandafter \@secondoftwo
 \fi
}%
\providecommand \@ifx [1]{%
 \ifx #1\expandafter \@firstoftwo
 \else \expandafter \@secondoftwo
 \fi
}%
\providecommand \natexlab [1]{#1}%
\providecommand \enquote  [1]{``#1''}%
\providecommand \bibnamefont  [1]{#1}%
\providecommand \bibfnamefont [1]{#1}%
\providecommand \citenamefont [1]{#1}%
\providecommand \href@noop [0]{\@secondoftwo}%
\providecommand \href [0]{\begingroup \@sanitize@url \@href}%
\providecommand \@href[1]{\@@startlink{#1}\@@href}%
\providecommand \@@href[1]{\endgroup#1\@@endlink}%
\providecommand \@sanitize@url [0]{\catcode `\\12\catcode `\$12\catcode
  `\&12\catcode `\#12\catcode `\^12\catcode `\_12\catcode `\%12\relax}%
\providecommand \@@startlink[1]{}%
\providecommand \@@endlink[0]{}%
\providecommand \url  [0]{\begingroup\@sanitize@url \@url }%
\providecommand \@url [1]{\endgroup\@href {#1}{\urlprefix }}%
\providecommand \urlprefix  [0]{URL }%
\providecommand \Eprint [0]{\href }%
\providecommand \doibase [0]{https://doi.org/}%
\providecommand \selectlanguage [0]{\@gobble}%
\providecommand \bibinfo  [0]{\@secondoftwo}%
\providecommand \bibfield  [0]{\@secondoftwo}%
\providecommand \translation [1]{[#1]}%
\providecommand \BibitemOpen [0]{}%
\providecommand \bibitemStop [0]{}%
\providecommand \bibitemNoStop [0]{.\EOS\space}%
\providecommand \EOS [0]{\spacefactor3000\relax}%
\providecommand \BibitemShut  [1]{\csname bibitem#1\endcsname}%
\let\auto@bib@innerbib\@empty
\bibitem [{\citenamefont {Anderson}(1958)}]{1958Anderson}%
  \BibitemOpen
  \bibfield  {author} {\bibinfo {author} {\bibfnamefont {P.}~\bibnamefont
  {Anderson}},\ }\bibfield  {title} {\bibinfo {title} {Absence of diffusion in
  certain random lattices},\ }\href@noop {} {\bibfield  {journal} {\bibinfo
  {journal} {Physical Review}\ }\textbf {\bibinfo {volume} {109}},\ \bibinfo
  {pages} {1492} (\bibinfo {year} {1958})}\BibitemShut {NoStop}%
\bibitem [{\citenamefont {Abrahams}\ \emph {et~al.}(1979)\citenamefont
  {Abrahams}, \citenamefont {Anderson}, \citenamefont {Licciardello},\ and\
  \citenamefont {Ramakrishnan}}]{1979Abrahams}%
  \BibitemOpen
  \bibfield  {author} {\bibinfo {author} {\bibfnamefont {E.}~\bibnamefont
  {Abrahams}}, \bibinfo {author} {\bibfnamefont {P.~W.}\ \bibnamefont
  {Anderson}}, \bibinfo {author} {\bibfnamefont {D.}~\bibnamefont
  {Licciardello}},\ and\ \bibinfo {author} {\bibfnamefont {T.~V.}\ \bibnamefont
  {Ramakrishnan}},\ }\bibfield  {title} {\bibinfo {title} {Scaling theory of
  localization: Absence of quantum diffusion in two dimensions},\ }\href@noop
  {} {\bibfield  {journal} {\bibinfo  {journal} {Phys. Rev. Lett.}\ }\textbf
  {\bibinfo {volume} {42}},\ \bibinfo {pages} {673} (\bibinfo {year}
  {1979})}\BibitemShut {NoStop}%
\bibitem [{\citenamefont {Page}(1993)}]{1993Page}%
  \BibitemOpen
  \bibfield  {author} {\bibinfo {author} {\bibfnamefont {D.}~\bibnamefont
  {Page}},\ }\bibfield  {title} {\bibinfo {title} {Average entropy of a
  subsystem},\ }\href@noop {} {\bibfield  {journal} {\bibinfo  {journal} {Phys.
  Rev. Lett.}\ }\textbf {\bibinfo {volume} {71}},\ \bibinfo {pages} {1291}
  (\bibinfo {year} {1993})}\BibitemShut {NoStop}%
\bibitem [{\citenamefont {Calabrese}\ and\ \citenamefont
  {Cardy}(2005)}]{2005Calabrese}%
  \BibitemOpen
  \bibfield  {author} {\bibinfo {author} {\bibfnamefont {P.}~\bibnamefont
  {Calabrese}}\ and\ \bibinfo {author} {\bibfnamefont {J.}~\bibnamefont
  {Cardy}},\ }\bibfield  {title} {\bibinfo {title} {Evolution of entanglement
  entropy in one-dimensional systems},\ }\href@noop {} {\bibfield  {journal}
  {\bibinfo  {journal} {Journal of Statistical Mechanics: Theory and
  Experiment}\ }\textbf {\bibinfo {volume} {04}},\ \bibinfo {pages} {P04010}
  (\bibinfo {year} {2005})}\BibitemShut {NoStop}%
\bibitem [{\citenamefont {Znidaric}\ \emph {et~al.}(2008)\citenamefont
  {Znidaric}, \citenamefont {Prosen},\ and\ \citenamefont
  {Prelovsek}}]{2008Znidaric}%
  \BibitemOpen
  \bibfield  {author} {\bibinfo {author} {\bibfnamefont {M.}~\bibnamefont
  {Znidaric}}, \bibinfo {author} {\bibfnamefont {T.}~\bibnamefont {Prosen}},\
  and\ \bibinfo {author} {\bibfnamefont {P.}~\bibnamefont {Prelovsek}},\
  }\bibfield  {title} {\bibinfo {title} {Many-body localization in the
  heisenberg xxz magnet in a random field},\ }\href@noop {} {\bibfield
  {journal} {\bibinfo  {journal} {Phys. Rev. B}\ }\textbf {\bibinfo {volume}
  {77}},\ \bibinfo {pages} {064426} (\bibinfo {year} {2008})}\BibitemShut
  {NoStop}%
\bibitem [{\citenamefont {Bardarson}\ \emph {et~al.}(2012)\citenamefont
  {Bardarson}, \citenamefont {Pollmann},\ and\ \citenamefont
  {Moore}}]{2012Bardarson}%
  \BibitemOpen
  \bibfield  {author} {\bibinfo {author} {\bibfnamefont {J.}~\bibnamefont
  {Bardarson}}, \bibinfo {author} {\bibfnamefont {F.}~\bibnamefont
  {Pollmann}},\ and\ \bibinfo {author} {\bibfnamefont {J.}~\bibnamefont
  {Moore}},\ }\bibfield  {title} {\bibinfo {title} {Unbounded growth of
  entanglement in models of many-body localization},\ }\href@noop {} {\bibfield
   {journal} {\bibinfo  {journal} {Phys. Rev. Lett.}\ }\textbf {\bibinfo
  {volume} {109}},\ \bibinfo {pages} {017202} (\bibinfo {year}
  {2012})}\BibitemShut {NoStop}%
\bibitem [{\citenamefont {Serbyn}\ \emph {et~al.}(2013)\citenamefont {Serbyn},
  \citenamefont {Papic},\ and\ \citenamefont {Abanin}}]{2013SerbynPRLJune}%
  \BibitemOpen
  \bibfield  {author} {\bibinfo {author} {\bibfnamefont {M.}~\bibnamefont
  {Serbyn}}, \bibinfo {author} {\bibfnamefont {Z.}~\bibnamefont {Papic}},\ and\
  \bibinfo {author} {\bibfnamefont {D.~A.}\ \bibnamefont {Abanin}},\ }\bibfield
   {title} {\bibinfo {title} {Universal slow growth of entanglement in
  interacting strongly disordered systems},\ }\href@noop {} {\bibfield
  {journal} {\bibinfo  {journal} {Phys. Rev. Lett.}\ }\textbf {\bibinfo
  {volume} {110}},\ \bibinfo {pages} {260601} (\bibinfo {year}
  {2013})}\BibitemShut {NoStop}%
\bibitem [{\citenamefont {Huse}\ \emph {et~al.}(2014)\citenamefont {Huse},
  \citenamefont {Nandkishore},\ and\ \citenamefont {Oganesyan}}]{2014Huse}%
  \BibitemOpen
  \bibfield  {author} {\bibinfo {author} {\bibfnamefont {D.}~\bibnamefont
  {Huse}}, \bibinfo {author} {\bibfnamefont {R.}~\bibnamefont {Nandkishore}},\
  and\ \bibinfo {author} {\bibfnamefont {V.}~\bibnamefont {Oganesyan}},\
  }\bibfield  {title} {\bibinfo {title} {Phenomenology of fully
  many-body-localized systems},\ }\href@noop {} {\bibfield  {journal} {\bibinfo
   {journal} {Phys. Rev. B}\ }\textbf {\bibinfo {volume} {90}},\ \bibinfo
  {pages} {174202} (\bibinfo {year} {2014})}\BibitemShut {NoStop}%
\bibitem [{\citenamefont {Nandkishore}\ and\ \citenamefont
  {Huse}(2015)}]{2015Nandkishore}%
  \BibitemOpen
  \bibfield  {author} {\bibinfo {author} {\bibfnamefont {R.}~\bibnamefont
  {Nandkishore}}\ and\ \bibinfo {author} {\bibfnamefont {D.}~\bibnamefont
  {Huse}},\ }\bibfield  {title} {\bibinfo {title} {Many-body localization and
  thermalization in quantum statistical mechanics},\ }\href@noop {} {\bibfield
  {journal} {\bibinfo  {journal} {Annual Review of Condensed Matter Physics}\
  }\textbf {\bibinfo {volume} {6}},\ \bibinfo {pages} {15} (\bibinfo {year}
  {2015})}\BibitemShut {NoStop}%
\bibitem [{\citenamefont {Abanin}\ \emph {et~al.}(2019)\citenamefont {Abanin},
  \citenamefont {Altman}, \citenamefont {Bloch},\ and\ \citenamefont
  {Serbyn}}]{2019Abanin}%
  \BibitemOpen
  \bibfield  {author} {\bibinfo {author} {\bibfnamefont {D.}~\bibnamefont
  {Abanin}}, \bibinfo {author} {\bibfnamefont {E.}~\bibnamefont {Altman}},
  \bibinfo {author} {\bibfnamefont {I.}~\bibnamefont {Bloch}},\ and\ \bibinfo
  {author} {\bibfnamefont {M.}~\bibnamefont {Serbyn}},\ }\bibfield  {title}
  {\bibinfo {title} {Ergodicity, entanglement and many-body localization},\
  }\href@noop {} {\bibfield  {journal} {\bibinfo  {journal} {Rev. Mod. Phys.}\
  }\textbf {\bibinfo {volume} {91}},\ \bibinfo {pages} {021001} (\bibinfo
  {year} {2019})}\BibitemShut {NoStop}%
\bibitem [{\citenamefont {Abanin}\ \emph {et~al.}(2021)\citenamefont {Abanin},
  \citenamefont {Bardarson}, \citenamefont {Tomasi}, \citenamefont
  {Gopalkrishnan}, \citenamefont {Khemani}, \citenamefont {Parameswaran},
  \citenamefont {Pollmann}, \citenamefont {Potter}, \citenamefont {Serbyn},\
  and\ \citenamefont {Vasseur}}]{2021Abanin}%
  \BibitemOpen
  \bibfield  {author} {\bibinfo {author} {\bibfnamefont {D.}~\bibnamefont
  {Abanin}}, \bibinfo {author} {\bibfnamefont {J.}~\bibnamefont {Bardarson}},
  \bibinfo {author} {\bibfnamefont {G.~D.}\ \bibnamefont {Tomasi}}, \bibinfo
  {author} {\bibfnamefont {S.}~\bibnamefont {Gopalkrishnan}}, \bibinfo {author}
  {\bibfnamefont {V.}~\bibnamefont {Khemani}}, \bibinfo {author} {\bibfnamefont
  {S.}~\bibnamefont {Parameswaran}}, \bibinfo {author} {\bibfnamefont
  {F.}~\bibnamefont {Pollmann}}, \bibinfo {author} {\bibfnamefont
  {A.}~\bibnamefont {Potter}}, \bibinfo {author} {\bibfnamefont
  {M.}~\bibnamefont {Serbyn}},\ and\ \bibinfo {author} {\bibfnamefont
  {R.}~\bibnamefont {Vasseur}},\ }\bibfield  {title} {\bibinfo {title}
  {Distinguishinglocalizationfromchaos: Distinguishing localization from chaos:
  Challenges in finite-size systems},\ }\href@noop {} {\bibfield  {journal}
  {\bibinfo  {journal} {Annals of Physics}\ }\textbf {\bibinfo {volume}
  {427}},\ \bibinfo {pages} {168415} (\bibinfo {year} {2021})}\BibitemShut
  {NoStop}%
\bibitem [{\citenamefont {Sierant}\ \emph {et~al.}(2025)\citenamefont
  {Sierant}, \citenamefont {Lewenstein}, \citenamefont {Scardicchio},
  \citenamefont {Vidmar},\ and\ \citenamefont {Zakrzewski}}]{2025Sierant}%
  \BibitemOpen
  \bibfield  {author} {\bibinfo {author} {\bibfnamefont {P.}~\bibnamefont
  {Sierant}}, \bibinfo {author} {\bibfnamefont {M.}~\bibnamefont {Lewenstein}},
  \bibinfo {author} {\bibfnamefont {A.}~\bibnamefont {Scardicchio}}, \bibinfo
  {author} {\bibfnamefont {L.}~\bibnamefont {Vidmar}},\ and\ \bibinfo {author}
  {\bibfnamefont {J.}~\bibnamefont {Zakrzewski}},\ }\bibfield  {title}
  {\bibinfo {title} {Many-body localization in the age of classical
  computing},\ }\href@noop {} {\bibfield  {journal} {\bibinfo  {journal} {Rep.
  Prog. Phys.}\ }\textbf {\bibinfo {volume} {88}},\ \bibinfo {pages} {026502}
  (\bibinfo {year} {2025})}\BibitemShut {NoStop}%
\bibitem [{\citenamefont {Prelovsek}\ \emph {et~al.}(2016)\citenamefont
  {Prelovsek}, \citenamefont {Barisic},\ and\ \citenamefont
  {Znidaric}}]{2016Prelovsek}%
  \BibitemOpen
  \bibfield  {author} {\bibinfo {author} {\bibfnamefont {P.}~\bibnamefont
  {Prelovsek}}, \bibinfo {author} {\bibfnamefont {O.}~\bibnamefont {Barisic}},\
  and\ \bibinfo {author} {\bibfnamefont {M.}~\bibnamefont {Znidaric}},\
  }\bibfield  {title} {\bibinfo {title} {Absence of full many-body localization
  in the disordered hubbard chain},\ }\href@noop {} {\bibfield  {journal}
  {\bibinfo  {journal} {Phys. Rev. B}\ }\textbf {\bibinfo {volume} {94}},\
  \bibinfo {pages} {241104(R)} (\bibinfo {year} {2016})}\BibitemShut {NoStop}%
\bibitem [{\citenamefont {Mierzejewski}\ \emph {et~al.}(2018)\citenamefont
  {Mierzejewski}, \citenamefont {Kozarzewski},\ and\ \citenamefont
  {Prelovsek}}]{2018Mierzejewski}%
  \BibitemOpen
  \bibfield  {author} {\bibinfo {author} {\bibfnamefont {M.}~\bibnamefont
  {Mierzejewski}}, \bibinfo {author} {\bibfnamefont {M.}~\bibnamefont
  {Kozarzewski}},\ and\ \bibinfo {author} {\bibfnamefont {P.}~\bibnamefont
  {Prelovsek}},\ }\bibfield  {title} {\bibinfo {title} {Counting local
  integrals of motion in disordered spinless-fermion and hubbard chains},\
  }\href@noop {} {\bibfield  {journal} {\bibinfo  {journal} {Phys. Rev. B}\
  }\textbf {\bibinfo {volume} {97}},\ \bibinfo {pages} {064204} (\bibinfo
  {year} {2018})}\BibitemShut {NoStop}%
\bibitem [{\citenamefont {Protopopov}\ and\ \citenamefont
  {Abanin}(2019)}]{2019Protopopov}%
  \BibitemOpen
  \bibfield  {author} {\bibinfo {author} {\bibfnamefont {I.}~\bibnamefont
  {Protopopov}}\ and\ \bibinfo {author} {\bibfnamefont {D.}~\bibnamefont
  {Abanin}},\ }\bibfield  {title} {\bibinfo {title} {Spin-mediated particle
  transport in the disordered hubbard model},\ }\href@noop {} {\bibfield
  {journal} {\bibinfo  {journal} {Phys. Rev. B}\ }\textbf {\bibinfo {volume}
  {99}},\ \bibinfo {pages} {115111} (\bibinfo {year} {2019})}\BibitemShut
  {NoStop}%
\bibitem [{\citenamefont {Leipner-Johns}\ and\ \citenamefont
  {Wortis}(2019)}]{2019Leipner-Johns}%
  \BibitemOpen
  \bibfield  {author} {\bibinfo {author} {\bibfnamefont {B.}~\bibnamefont
  {Leipner-Johns}}\ and\ \bibinfo {author} {\bibfnamefont {R.}~\bibnamefont
  {Wortis}},\ }\bibfield  {title} {\bibinfo {title} {Charge- and spin-specific
  local integrals of motion in a disordered hubbard model},\ }\href@noop {}
  {\bibfield  {journal} {\bibinfo  {journal} {Phys. Rev. B}\ }\textbf {\bibinfo
  {volume} {100}},\ \bibinfo {pages} {125132} (\bibinfo {year}
  {2019})}\BibitemShut {NoStop}%
\bibitem [{\citenamefont {Thomson}(2023)}]{2023Thomson}%
  \BibitemOpen
  \bibfield  {author} {\bibinfo {author} {\bibfnamefont {S.}~\bibnamefont
  {Thomson}},\ }\bibfield  {title} {\bibinfo {title} {Disorder-induced
  spin-charge separation in the one-dimensional hubbard model},\ }\href@noop {}
  {\bibfield  {journal} {\bibinfo  {journal} {Phys. Rev. B}\ }\textbf {\bibinfo
  {volume} {107}},\ \bibinfo {pages} {L180201} (\bibinfo {year}
  {2023})}\BibitemShut {NoStop}%
\bibitem [{\citenamefont {Yu}\ \emph {et~al.}(2018)\citenamefont {Yu},
  \citenamefont {Luo},\ and\ \citenamefont {Clark}}]{2018Yu}%
  \BibitemOpen
  \bibfield  {author} {\bibinfo {author} {\bibfnamefont {X.}~\bibnamefont
  {Yu}}, \bibinfo {author} {\bibfnamefont {D.}~\bibnamefont {Luo}},\ and\
  \bibinfo {author} {\bibfnamefont {B.}~\bibnamefont {Clark}},\ }\bibfield
  {title} {\bibinfo {title} {Beyond many-body localized states in a
  spin-disordered hubbard model},\ }\href@noop {} {\bibfield  {journal}
  {\bibinfo  {journal} {Phys. Rev. B}\ }\textbf {\bibinfo {volume} {98}},\
  \bibinfo {pages} {115106} (\bibinfo {year} {2018})}\BibitemShut {NoStop}%
\bibitem [{\citenamefont {Lev}\ \emph {et~al.}(2016)\citenamefont {Lev},
  \citenamefont {Reichman},\ and\ \citenamefont {Sagi}}]{2016BarLevPRB}%
  \BibitemOpen
  \bibfield  {author} {\bibinfo {author} {\bibfnamefont {Y.~B.}\ \bibnamefont
  {Lev}}, \bibinfo {author} {\bibfnamefont {D.}~\bibnamefont {Reichman}},\ and\
  \bibinfo {author} {\bibfnamefont {Y.}~\bibnamefont {Sagi}},\ }\bibfield
  {title} {\bibinfo {title} {Many-body localization in system with a completely
  delocalized single-particle spectrum},\ }\href@noop {} {\bibfield  {journal}
  {\bibinfo  {journal} {Phys. Rev. B}\ }\textbf {\bibinfo {volume} {94}},\
  \bibinfo {pages} {201116} (\bibinfo {year} {2016})}\BibitemShut {NoStop}%
\bibitem [{\citenamefont {Potter}\ and\ \citenamefont
  {Vasseur}(2016)}]{2016Potter}%
  \BibitemOpen
  \bibfield  {author} {\bibinfo {author} {\bibfnamefont {A.}~\bibnamefont
  {Potter}}\ and\ \bibinfo {author} {\bibfnamefont {R.}~\bibnamefont
  {Vasseur}},\ }\bibfield  {title} {\bibinfo {title} {Symmetry constraints on
  many-body localization},\ }\href@noop {} {\bibfield  {journal} {\bibinfo
  {journal} {Phys. Rev. B}\ }\textbf {\bibinfo {volume} {94}},\ \bibinfo
  {pages} {224206} (\bibinfo {year} {2016})}\BibitemShut {NoStop}%
\bibitem [{\citenamefont {Protopopov}\ \emph {et~al.}(2017)\citenamefont
  {Protopopov}, \citenamefont {Ho},\ and\ \citenamefont
  {Abanin}}]{2017Protopopov}%
  \BibitemOpen
  \bibfield  {author} {\bibinfo {author} {\bibfnamefont {I.}~\bibnamefont
  {Protopopov}}, \bibinfo {author} {\bibfnamefont {W.~W.}\ \bibnamefont {Ho}},\
  and\ \bibinfo {author} {\bibfnamefont {D.}~\bibnamefont {Abanin}},\
  }\bibfield  {title} {\bibinfo {title} {Effect of su(2) symmetry on many-body
  localization and thermalization},\ }\href@noop {} {\bibfield  {journal}
  {\bibinfo  {journal} {Phys. Rev. B}\ }\textbf {\bibinfo {volume} {96}},\
  \bibinfo {pages} {041122(R)} (\bibinfo {year} {2017})}\BibitemShut {NoStop}%
\bibitem [{\citenamefont {Kozarzewski}\ \emph {et~al.}(2018)\citenamefont
  {Kozarzewski}, \citenamefont {Prelovsek},\ and\ \citenamefont
  {Mierzejewski}}]{2018Kozarzewski}%
  \BibitemOpen
  \bibfield  {author} {\bibinfo {author} {\bibfnamefont {M.}~\bibnamefont
  {Kozarzewski}}, \bibinfo {author} {\bibfnamefont {P.}~\bibnamefont
  {Prelovsek}},\ and\ \bibinfo {author} {\bibfnamefont {M.}~\bibnamefont
  {Mierzejewski}},\ }\bibfield  {title} {\bibinfo {title} {Spin subdiffusion in
  disordered hubbard chain},\ }\href@noop {} {\bibfield  {journal} {\bibinfo
  {journal} {Phys. Rev. Lett.}\ }\textbf {\bibinfo {volume} {120}},\ \bibinfo
  {pages} {246602} (\bibinfo {year} {2018})}\BibitemShut {NoStop}%
\bibitem [{\citenamefont {Bonca}\ and\ \citenamefont
  {Mierzejewski}(2022)}]{2022Bonca}%
  \BibitemOpen
  \bibfield  {author} {\bibinfo {author} {\bibfnamefont {J.}~\bibnamefont
  {Bonca}}\ and\ \bibinfo {author} {\bibfnamefont {M.}~\bibnamefont
  {Mierzejewski}},\ }\bibfield  {title} {\bibinfo {title} {Relaxation
  mechanisms in a disordered system with poisson-level statistics},\
  }\href@noop {} {\bibfield  {journal} {\bibinfo  {journal} {Phys. Rev. B}\
  }\textbf {\bibinfo {volume} {105}},\ \bibinfo {pages} {155146} (\bibinfo
  {year} {2022})}\BibitemShut {NoStop}%
\bibitem [{\citenamefont {Iadecola}\ and\ \citenamefont
  {Znidaric}(2019)}]{2019Iadecola}%
  \BibitemOpen
  \bibfield  {author} {\bibinfo {author} {\bibfnamefont {T.}~\bibnamefont
  {Iadecola}}\ and\ \bibinfo {author} {\bibfnamefont {M.}~\bibnamefont
  {Znidaric}},\ }\bibfield  {title} {\bibinfo {title} {Exact localized and
  ballistic eigenstates in disordered chaotic spin ladders and the
  fermi-hubbard model},\ }\href@noop {} {\bibfield  {journal} {\bibinfo
  {journal} {Phys. Rev. Lett.}\ }\textbf {\bibinfo {volume} {123}},\ \bibinfo
  {pages} {036403} (\bibinfo {year} {2019})}\BibitemShut {NoStop}%
\bibitem [{\citenamefont {Bahovadinov}\ \emph {et~al.}(2022)\citenamefont
  {Bahovadinov}, \citenamefont {Kurlov}, \citenamefont {Altshuler},\ and\
  \citenamefont {Shlyapnikov}}]{2022Bahovadinov}%
  \BibitemOpen
  \bibfield  {author} {\bibinfo {author} {\bibfnamefont {M.}~\bibnamefont
  {Bahovadinov}}, \bibinfo {author} {\bibfnamefont {D.}~\bibnamefont {Kurlov}},
  \bibinfo {author} {\bibfnamefont {B.}~\bibnamefont {Altshuler}},\ and\
  \bibinfo {author} {\bibfnamefont {G.}~\bibnamefont {Shlyapnikov}},\
  }\bibfield  {title} {\bibinfo {title} {Many-body localization of 1d
  disordered impenetrable two-component fermions},\ }\href@noop {} {\bibfield
  {journal} {\bibinfo  {journal} {The European Physics Journal D}\ }\textbf
  {\bibinfo {volume} {76}},\ \bibinfo {pages} {116} (\bibinfo {year}
  {2022})}\BibitemShut {NoStop}%
\bibitem [{\citenamefont {Sroda}\ \emph {et~al.}(2019)\citenamefont {Sroda},
  \citenamefont {Prelovsek},\ and\ \citenamefont {Mierzejewski}}]{2019Sroda}%
  \BibitemOpen
  \bibfield  {author} {\bibinfo {author} {\bibfnamefont {M.}~\bibnamefont
  {Sroda}}, \bibinfo {author} {\bibfnamefont {P.}~\bibnamefont {Prelovsek}},\
  and\ \bibinfo {author} {\bibfnamefont {M.}~\bibnamefont {Mierzejewski}},\
  }\bibfield  {title} {\bibinfo {title} {Instability of subdiffusive spin
  dynamics in strongly disordered hubbard chain},\ }\href@noop {} {\bibfield
  {journal} {\bibinfo  {journal} {Phys. Rev. B}\ }\textbf {\bibinfo {volume}
  {99}},\ \bibinfo {pages} {121110(R)} (\bibinfo {year} {2019})}\BibitemShut
  {NoStop}%
\bibitem [{\citenamefont {Pandey}\ \emph {et~al.}(2020)\citenamefont {Pandey},
  \citenamefont {Dagotto},\ and\ \citenamefont {Pati}}]{2020Pandey}%
  \BibitemOpen
  \bibfield  {author} {\bibinfo {author} {\bibfnamefont {B.}~\bibnamefont
  {Pandey}}, \bibinfo {author} {\bibfnamefont {E.}~\bibnamefont {Dagotto}},\
  and\ \bibinfo {author} {\bibfnamefont {S.~K.}\ \bibnamefont {Pati}},\
  }\bibfield  {title} {\bibinfo {title} {Quench dynamics of two-component
  dipolar fermions subject to a quasiperiodic potential},\ }\href@noop {}
  {\bibfield  {journal} {\bibinfo  {journal} {Phys. Rev. B}\ }\textbf {\bibinfo
  {volume} {102}},\ \bibinfo {pages} {214302} (\bibinfo {year}
  {2020})}\BibitemShut {NoStop}%
\bibitem [{\citenamefont {Zakrzewski}\ and\ \citenamefont
  {Delande}(2018)}]{2018Zakrzewski}%
  \BibitemOpen
  \bibfield  {author} {\bibinfo {author} {\bibfnamefont {J.}~\bibnamefont
  {Zakrzewski}}\ and\ \bibinfo {author} {\bibfnamefont {D.}~\bibnamefont
  {Delande}},\ }\bibfield  {title} {\bibinfo {title} {Spin-charge separation
  and many-body localization},\ }\href@noop {} {\bibfield  {journal} {\bibinfo
  {journal} {Phys. Rev. B}\ }\textbf {\bibinfo {volume} {98}},\ \bibinfo
  {pages} {014203} (\bibinfo {year} {2018})}\BibitemShut {NoStop}%
\bibitem [{\citenamefont {Islam}\ \emph {et~al.}(2015)\citenamefont {Islam},
  \citenamefont {Ma}, \citenamefont {Preiss}, \citenamefont {Tai},
  \citenamefont {Lukin}, \citenamefont {Rispoli},\ and\ \citenamefont
  {Greiner}}]{2015Islam}%
  \BibitemOpen
  \bibfield  {author} {\bibinfo {author} {\bibfnamefont {R.}~\bibnamefont
  {Islam}}, \bibinfo {author} {\bibfnamefont {R.}~\bibnamefont {Ma}}, \bibinfo
  {author} {\bibfnamefont {P.}~\bibnamefont {Preiss}}, \bibinfo {author}
  {\bibfnamefont {M.}~\bibnamefont {Tai}}, \bibinfo {author} {\bibfnamefont
  {A.}~\bibnamefont {Lukin}}, \bibinfo {author} {\bibfnamefont
  {M.}~\bibnamefont {Rispoli}},\ and\ \bibinfo {author} {\bibfnamefont
  {M.}~\bibnamefont {Greiner}},\ }\bibfield  {title} {\bibinfo {title}
  {Measuring entanglement entropy in a quantum many-body system},\ }\href@noop
  {} {\bibfield  {journal} {\bibinfo  {journal} {Nature}\ }\textbf {\bibinfo
  {volume} {528}},\ \bibinfo {pages} {77} (\bibinfo {year} {2015})}\BibitemShut
  {NoStop}%
\bibitem [{\citenamefont {Lukin}\ \emph {et~al.}(2019)\citenamefont {Lukin},
  \citenamefont {Rispoli}, \citenamefont {Schittko}, \citenamefont {Tai},
  \citenamefont {Kaufman}, \citenamefont {Choi}, \citenamefont {Khemani},
  \citenamefont {Leonard},\ and\ \citenamefont {Greiner}}]{2019Lukin}%
  \BibitemOpen
  \bibfield  {author} {\bibinfo {author} {\bibfnamefont {A.}~\bibnamefont
  {Lukin}}, \bibinfo {author} {\bibfnamefont {M.}~\bibnamefont {Rispoli}},
  \bibinfo {author} {\bibfnamefont {R.}~\bibnamefont {Schittko}}, \bibinfo
  {author} {\bibfnamefont {M.~E.}\ \bibnamefont {Tai}}, \bibinfo {author}
  {\bibfnamefont {A.~M.}\ \bibnamefont {Kaufman}}, \bibinfo {author}
  {\bibfnamefont {S.}~\bibnamefont {Choi}}, \bibinfo {author} {\bibfnamefont
  {V.}~\bibnamefont {Khemani}}, \bibinfo {author} {\bibfnamefont
  {J.}~\bibnamefont {Leonard}},\ and\ \bibinfo {author} {\bibfnamefont
  {M.}~\bibnamefont {Greiner}},\ }\bibfield  {title} {\bibinfo {title} {Probing
  entanglement in a many-body-localized system},\ }\href@noop {} {\bibfield
  {journal} {\bibinfo  {journal} {Science}\ }\textbf {\bibinfo {volume}
  {364}},\ \bibinfo {pages} {256} (\bibinfo {year} {2019})}\BibitemShut
  {NoStop}%
\bibitem [{\citenamefont {Leonard}\ \emph {et~al.}(2023)\citenamefont
  {Leonard}, \citenamefont {Kim}, \citenamefont {Rispoli}, \citenamefont
  {Lukin}, \citenamefont {Schittko}, \citenamefont {Kwan}, \citenamefont
  {Demler}, \citenamefont {Sels},\ and\ \citenamefont {Greiner}}]{2023Leonard}%
  \BibitemOpen
  \bibfield  {author} {\bibinfo {author} {\bibfnamefont {J.}~\bibnamefont
  {Leonard}}, \bibinfo {author} {\bibfnamefont {S.}~\bibnamefont {Kim}},
  \bibinfo {author} {\bibfnamefont {M.}~\bibnamefont {Rispoli}}, \bibinfo
  {author} {\bibfnamefont {A.}~\bibnamefont {Lukin}}, \bibinfo {author}
  {\bibfnamefont {R.}~\bibnamefont {Schittko}}, \bibinfo {author}
  {\bibfnamefont {J.}~\bibnamefont {Kwan}}, \bibinfo {author} {\bibfnamefont
  {E.}~\bibnamefont {Demler}}, \bibinfo {author} {\bibfnamefont
  {D.}~\bibnamefont {Sels}},\ and\ \bibinfo {author} {\bibfnamefont
  {M.}~\bibnamefont {Greiner}},\ }\bibfield  {title} {\bibinfo {title} {Probing
  the onset of quantum avalanches in a many-body localized system},\
  }\href@noop {} {\bibfield  {journal} {\bibinfo  {journal} {Nature Physics}\
  }\textbf {\bibinfo {volume} {19}},\ \bibinfo {pages} {481} (\bibinfo {year}
  {2023})}\BibitemShut {NoStop}%
\bibitem [{\citenamefont {Chiaro}\ \emph {et~al.}(2022)\citenamefont {Chiaro},
  \citenamefont {Neill}, \citenamefont {Bohrdt}, \citenamefont {Filippone},
  \citenamefont {Arute}, \citenamefont {Arya},\ and\ \citenamefont {{\it et
  al}}}]{2022Chiaro}%
  \BibitemOpen
  \bibfield  {author} {\bibinfo {author} {\bibfnamefont {B.}~\bibnamefont
  {Chiaro}}, \bibinfo {author} {\bibfnamefont {C.}~\bibnamefont {Neill}},
  \bibinfo {author} {\bibfnamefont {A.}~\bibnamefont {Bohrdt}}, \bibinfo
  {author} {\bibfnamefont {M.}~\bibnamefont {Filippone}}, \bibinfo {author}
  {\bibfnamefont {F.}~\bibnamefont {Arute}}, \bibinfo {author} {\bibfnamefont
  {K.}~\bibnamefont {Arya}},\ and\ \bibinfo {author} {\bibnamefont {{\it et
  al}}},\ }\bibfield  {title} {\bibinfo {title} {Direct measurement of nonlocal
  interactions in the many-body localized phase},\ }\href@noop {} {\bibfield
  {journal} {\bibinfo  {journal} {Phys. Rev. Research}\ }\textbf {\bibinfo
  {volume} {4}},\ \bibinfo {pages} {013148} (\bibinfo {year}
  {2022})}\BibitemShut {NoStop}%
\end{thebibliography}

%

\end{document}